\documentclass[runningheads]{llncs}
\usepackage[T1]{fontenc}
\usepackage{graphicx,verbatim}
\begin{document}
\title{TReND: \underline{T}ransformer derived features and \underline{Re}gularized \underline{N}MF for neonatal functional network \underline{D}elineation}
%
%% Removed for anonymized MICCAI 2025 submission
\author{Sovesh Mohapatra\inst{1,2} \and Minhui Ouyang\inst{2,3} \and Shufang Tan\inst{2,4} \and Jianlin Guo\inst{2,5} \and Lianglong Sun\inst{6} \and Yong He\inst{6} \and Hao Huang\inst{2,3}}
\authorrunning{Sovesh Mohapatra et al.}
\institute{Department of Bioengineering, School of Engineering and Applied Science, University of Pennsylvania, Philadelphia, PA, United States \and Department of Radiology, Children's Hospital of Philadelphia, Philadelphia, PA, United States \and Department of Radiology, Perelman School of Medicine, University of Pennsylvania, Philadelphia, PA, United States \and Graduate School of Education, Peking University, Beijing, China \and Department of Radiology, Beijing Children's  Hospital, National Center for Children's health, Beijing, China \and State Key Laboratory of Cognitive Neuroscience and Learning, Beijing Normal University, Beijing, China}

% \author{Anonymized Authors}  %% Added for anonymized MICCAI 2025 submission
% \authorrunning{Anonymized Author et al.}
% \institute{Anonymized Affiliations \\
%     \email{email@anonymized.com}}

\maketitle              % typeset the header of the contribution
\begin{abstract}
Precise parcellation of functional networks (FNs) of early developing human brain is the fundamental basis for identifying biomarker of developmental disorders and understanding functional development. Resting-state fMRI (rs-fMRI) enables in vivo exploration of functional changes, but adult FN parcellations cannot be directly applied to the neonates due to incomplete network maturation. No standardized neonatal functional atlas is currently available.  To solve this fundamental issue, we propose TReND, a novel and fully automated self-supervised transformer-autoencoder framework that integrates regularized nonnegative matrix factorization (RNMF) to unveil the FNs in neonates. TReND effectively disentangles spatiotemporal features in voxel-wise rs-fMRI data. The framework integrates confidence-adaptive masks into transformer self-attention layers to mitigate noise influence. A self supervised decoder acts as a regulator to refine the encoder's latent embeddings, which serve as reliable temporal features. For spatial coherence, we incorporate brain surface-based geodesic distances as spatial encodings along with functional connectivity from temporal features. The TReND clustering approach processes these features under sparsity and smoothness constraints, producing robust and biologically plausible parcellations. We extensively validated our TReND framework on three different rs-fMRI datasets: simulated, dHCP and HCP-YA against comparable traditional feature extraction and clustering techniques. Our results demonstrated the superiority of the TReND framework in the delineation of neonate FNs with significantly better spatial contiguity and functional homogeneity. Collectively, we established TReND, a novel and robust framework, for neonatal FN delineation. TReND-derived neonatal FNs could serve as a neonatal functional atlas for perinatal populations in health and disease.

% Authors must provide keywords and are not allowed to remove this Keyword section.

\end{abstract}
\section{Introduction}
Neuroscientists have long attempted to subdivide the human brain into a mesh of anatomically and functionally distinct, contiguous regions \cite{ref1,ref2,ref3,ref4,ref5,ref25}. This challenge become particularly complex in the neonatal brain, where functional organization differs markedly from that of adults. During the third trimester, the neonatal brain undergoes a critical phase of enhanced functional segregation, largely driven by the rapid development of functional connectivity (FC) and the formation of hubs in primary regions \cite{ref6,ref7,ref8}. This period of growth and organization leads to the emergence of functionally segregated networks, revealing the underlying principles that shape both healthy and diseased brain states \cite{ref9,ref10,ref11}. However, achieving accurate and reliable parcellation of specific functional networks (FNs) in newborns presents unique challenges. The combined effects of rapid functional segregation, low imaging quality, and the absence of established functional atlases as prior reference knowledge complicate this process \cite{ref12,ref13}.
\begin{figure}
\includegraphics[width=\textwidth]{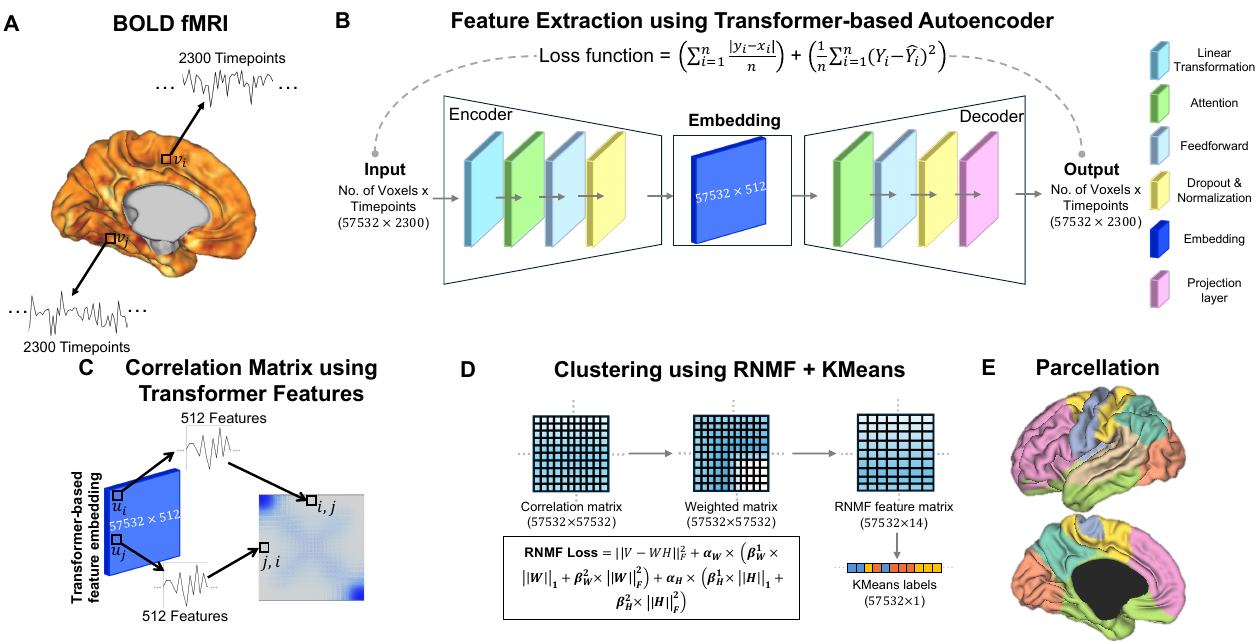}
\caption{Schematic representation of end-to-end functional parcellation framework. \textbf{A.} Processed Blood Oxygenation Level Dependent (BOLD) fMRI data. \textbf{B.} Feature extraction using transformer-based autoencoder to convert BOLD signals into low-dimensional embeddings. \textbf{C.} Correlation matrix from the transformer-based feature embeddings. \textbf{D.} Clustering using RNMF $+$ KMeans are performed to generate parcellation of different regions. \textbf{E.} Parcellation shows different regions based on correlation matrix.} \label{fig1}
\end{figure}

In this study, we developed the \underline{T}ransformer derived \underline{Re}gularized \underline{N}MF for neonatal FN \underline{D}elineation (TReND) framework. This approach integrates a novel transformer-based autoencoder architecture to extract features from rs-fMRI signals with an innovative regularized nonnegative matrix factorization (RNMF) clustering algorithm for the delineation of FNs. 
\section{Methods}
\subsection{Data description}
The study included three datasets. The simulated dataset contained 200 2D images ($100\times100$ pixels, 2300 timepoints) with 15 FNs \cite{ref14}. The developing human connectome project (dHCP) included 300 term neonates (mean scan age: 41.16 weeks) \cite{ref15}. All data were preprocessed and registered to the 40 week neonatal surface ($\sim$32K vertices per hemisphere, 2300 timepoints) \cite{ref16}. HCP Young Adult (HCP-YA) comprised of 200 randomly selected subjects and all were preprocessed and registered to fsLR surface with 1200 timepoints \cite{ref17,ref18}.
% Additionally, an in-house cohort consisted of 19 term neonates (mean scan age: 39.48 weeks).
\subsection{Overview of TReND framework}
TReND is an automated self-supervised transformer-based autoencoder framework that incorporates confidence-adaptive masks to extract salient temporal features and employs geodesic-guided RNMF-KMeans clustering for robust neonatal FN delineation. (Fig ~\ref{fig1}).
\subsubsection{Transformer based autoencoder with confidence-adaptive mask for temporal feature extraction.} The architecture is designed to capture long-range temporal dependencies while mitigating the over-influence of noise. The autoencoder consists of two main components: an encoder \(f_\theta(\cdot)\) that maps the input fMRI signal (Fig ~\ref{fig1}A) to a latent representation and a decoder acting as a self-regulator \(g_\phi(\cdot)\) that reconstructs the original signal from the latent representation. The encoder is built by stacking multiple transformer layers that capture long-range temporal dependencies via self-attention (Fig~\ref{fig1}B and ~\ref{fig2}A-\textit{left panel}) \cite{ref19}. The attention operation is being modified to include a confidence-adaptive mask \textbf{M}:
\begin{equation}
\mathrm{Attention}(Q,K,V) = \mathrm{softmax} \left( \frac{QK^\top}{\sqrt{d_k}} + M \right) V
\end{equation}
Here, \textit{Q}, \textit{K}, and \textit{V} represent the query, key, and value matrices obtained from linear projections of the input $d_k$ (e.g. 512) is the dimensionality of the key vectors, and the confidence mask \textit{M} modulates the contribution of each token based on its reliability. 

A central innovation of our approach is the ConfidenceModule: a two-layer multilayer perceptron (MLP) that estimates the reliability of each embedded token. Given an embedded sequence $x_{emb} \in R^{T \times B \times d_{k}}$ (where $B$ is the batch size, $T$ is the number of time points), the module computes a scalar confidence score for each token. These scores, which range between 0 and 1, are transformed into log domain values and combined pairwise to form the additive confidence mask.
\begin{enumerate}
    \item Token-wise confidence estimation: Each token is passed through the ConfidenceModule, which applies a ReLU activation followed by a sigmoid function to produce a confidence score that reflects the reliability of the token.
    \item Mask formation: The log-transformed confidence scores are summed pairwise to produce a mask \textbf{M} such that for tokens \textit{i} and \textit{j}, the mask entry is given by:
\begin{equation}
    \mathrm{M}_{ij} = \log (c_i + \epsilon) + \log (c_j + \epsilon)
\end{equation}
where $c_{i}$ and $c_{j}$ denote the confidence scores for tokens \textit{i} and \textit{j} respectively, and $\epsilon$ is a small constant to avoid numerical instability.
\end{enumerate}
This confidence-adaptive mechanism implicitly detects and down-weights over-influencing tokens during the attention computation, ensuring that the latent representations are dominated by reliable and discriminative features. 

The transformer-based autoencoder, trained in a self-supervised manner with a reconstruction objective, inherently regularizes the latent space, guiding the encoder to extract the most salient temporal features.

\subsubsection{Feature extraction loss.} To balance the trade-off between bias and variance in the reconstruction, we employ a composite loss function that combines Mean Squared Error (MSE) and the Mean Absolute Error (MAE):
\begin{equation}
    \mathcal{L} = \lambda_{\mathrm{MSE}} \cdot \mathcal{L}_{\mathrm{MSE}} + \lambda_{\mathrm{MAE}} \cdot \mathcal{L}_{\mathrm{MAE}}
\end{equation}
\begin{equation}
\mathcal{L}_{\mathrm{MSE}} = \frac{1}{N} \sum_{i=1}^{N} (x_i - \hat{x}_i)^2 \quad
\mathcal{L}_{\mathrm{MAE}} = \frac{1}{N} \sum_{i=1}^{N} |x_i - \hat{x}_i|
\end{equation}
where N is the is the total number of data points across time and voxels, and $\lambda_{MSE}$ and $\lambda_{MAE}$ are hyperparameters that adjust the relative contribution of each loss component.
\subsubsection{Regularized non-negative matrix factorization with KMeans for clustering.} 
\begin{figure}
\includegraphics[width=\textwidth]{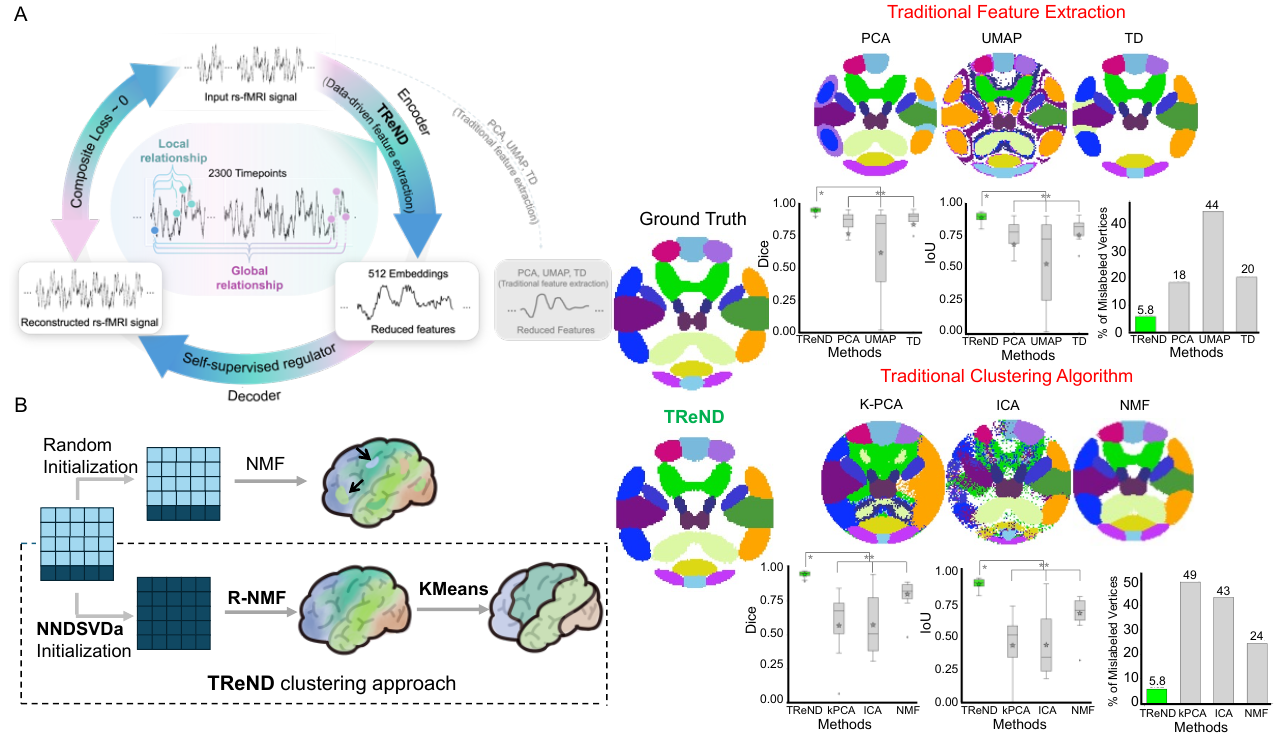}
\caption{Performance evaluation of TReND versus traditional parcellation methods on simulated data. \textbf{A.} \textit{Left panel}: The cartoon illustrates how TReND extracts features and preserves more fine-grained details. \textit{Right panel}:  Comparison of feature extraction techniques: Principal component analysis (PCA), Uniform manifold approximation and projection (UMAP), Tensor decomposition (TD), and TReND. \textbf{B.} \textit{Left panel}: The cartoon shows how TReND produces spatially coherent clusters. \textit{Right panel}: Comparison of clustering methods: kernel-PCA (kPCA), Independent component analysis (ICA), NMF, and TReND.} \label{fig2}
\end{figure}

To robustly parcellate neonatal FNs, we propose a clustering approach that synergistically integrates an advanced initialization strategy with regularization parameters \cite{ref20}. Our method begins with the construction of the data matrix (\textbf{V}) which is derived by integrating FC matrix (Fig~\ref{fig1}C and~\ref{fig2}B-\textit{left panel}) obtained via temporal features with spatial encodings derived from brain surface-based geodesic distances. This constructed data matrix is then decomposed using RNMF.

As NMF algorithms are well known for their sensitivity to initialization; random initialization often leads to suboptimal or unstable solutions. To mitigate this, we employ the non-negative double singular value decomposition with averaging (NNDSVDa) to initialize \textbf{W} (basis matrix) and \textbf{H} (encoding matrix) \cite{ref21}. This approach provides a more reliable starting point than random initialization, helping to position the algorithm in a favorable region of the solution space. 

\subsubsection{Clustering loss.} To extract functionally and spatially coherent regions, we introduce a loss function that both minimizes the reconstruction error and incorporates regularization terms to promote sparsity and smoothness in the decomposed factors:
\begin{equation}
    \begin{array}{c}
        \mathcal{L} = \|\mathrm{V} - \mathrm{W} \mathrm{H}\|^2_{F} 
        + \alpha_{\mathrm{W}} \times 
        \big( \beta^{1}_{\mathrm{W}} \times \|\mathrm{W}\|_{1} 
        + \beta^{2}_{\mathrm{W}} \times \|\mathrm{W}\|^2_{F} \big) \\[5pt]  
        + \alpha_{\mathrm{H}} \times 
        \big( \beta^{1}_{\mathrm{H}} \times \|\mathrm{H}\|_{1} 
        + \beta^{2}_{\mathrm{H}} \times \|\mathrm{H}\|^2_{F} \big)
    \end{array}
\end{equation}
where $\|.\|_{F}$ denotes the Frobenius norm, ensuring that the overall reconstruction error is minimized. $\|.\|_{1}$ promotes sparsity in the factors, essential for isolating brain networks. The parameters $\alpha_{W}$ and $\alpha_{H}$ control the overall regularization strength for the basis and encoding matrices, respectively. $\beta^1_{W,H}$ and $\beta^2_{W,H}$ balance the respective weighted contributions of the sparsity and smoothness.

After factorizing $\textbf{V} \approx \textbf{W$\times$H}$, we obtain a latent representation of each brain voxel in \textbf{W}. To define final parcels, we apply KMeans on \textbf{W} (rows of \textbf{W} correspond to voxels, columns correspond to latent features) (Fig ~\ref{fig1}D). This additional clustering step refines each row’s component assignments into discrete labels, thereby producing coherent, biologically plausible neonatal functional brain parcellations (Fig ~\ref{fig1}E).
\subsection{Evaluation of TReND framework}
Given the absence of well-established functional atlases for neonates, we evaluated TReND using a simulated dataset with 15 FNs as ground truth. We then applied to the dHCP cohort with 300 term neonates. To find optimal starting points for delineation of FNs, we assessed cluster stability by splitting the $\sim$64K vertices into two groups, clustering them independently, and using learned parameters from one to predict clusters in the other \cite{ref22}. The agreement between predicted and derived clusters reflects the stability. Subsequently, the spatial confidence map is calculated using the silhouette measure, which quantifies correlation-based similarity to the assigned network versus the nearest alternative \cite{ref23}. The validation is further performed via bootstrap analysis across 100 combinations using discovery and replication subsets (150 subjects each) of the dHCP cohort. TReND's applicability is also tested on the HCP-YA dataset using the Yeo 7-Network atlas \cite{ref3}. 
% and a smaller in-house neonatal cohort with distinct imaging parameters and demographics.
\begin{figure}
\includegraphics[width=\textwidth]{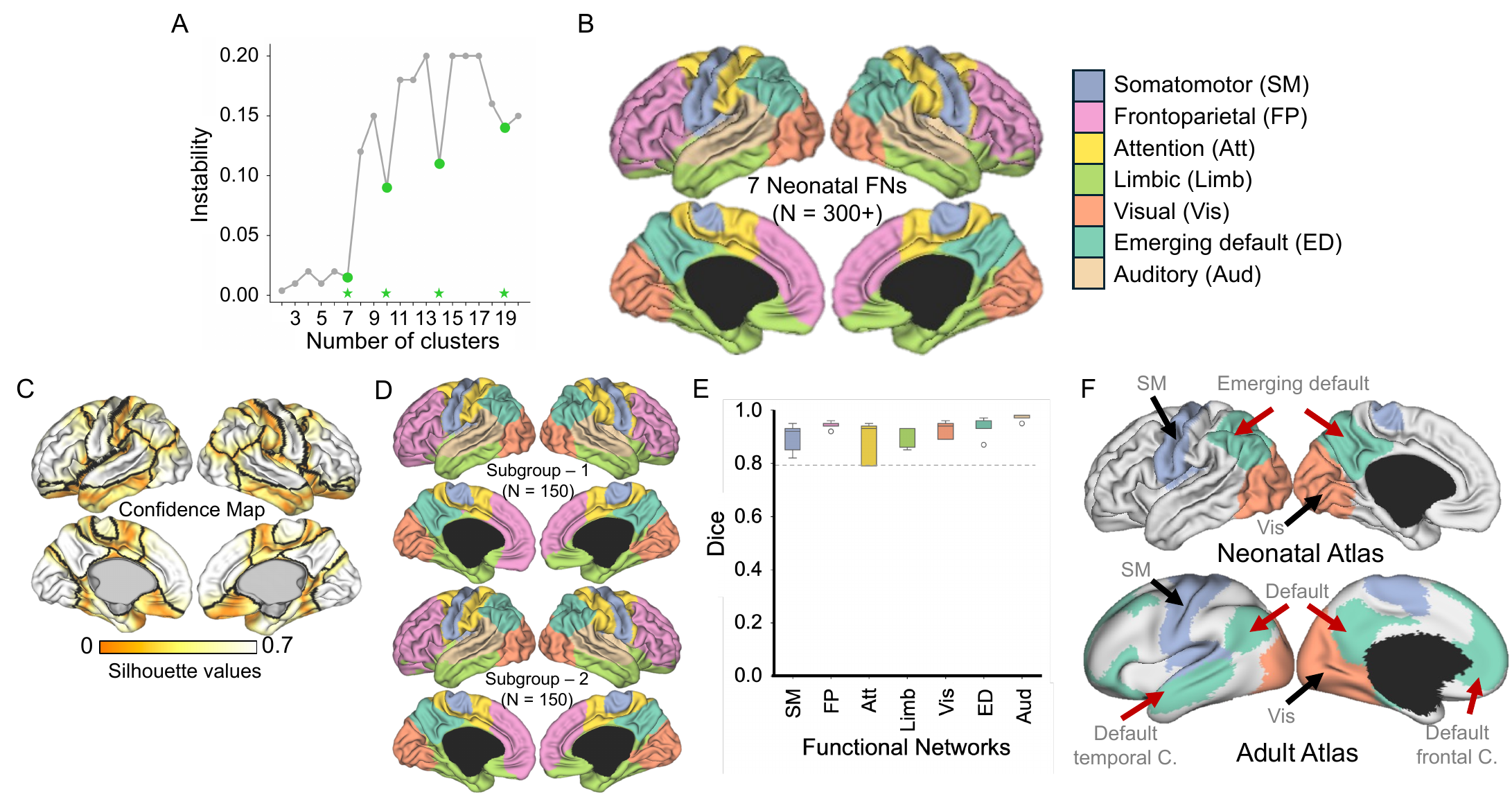}
\caption{Coarse parcellation of 7 functional networks in neonates derived from a 300-subject dHCP cohort. \textbf{A.} Stability analysis of the clustering algorithm identifies 7 and 19 networks as robust estimates. \textbf{B.} Neonatal cortical parcellation of 7 FNs. \textbf{C.} Confidence map representing the reliability of the 7 neonatal FNs parcellation. \textbf{D.} Discovery and replication of the 7 neonatal FNs cortical parcellation. \textbf{E.} Dice score evaluation FN-wise. \textbf{F.} Comparison of primary and higher-order FNs in TReND-derived neonate and adult atlas \cite{ref3}. (Abbreviation: C.: Component)}\label{fig3}
\end{figure}
\section{Results}
\subsection{Validation on simulated dataset}
Before applying TReND to real-world datasets, we first validated its ability to parcellate FNs using a controlled simulated dataset. The robustness of our transformer-based autoencoder for feature extraction was tested by keeping the RNMF-KMeans clustering consistent against other conventional mathematically constrained methods, PCA, UMAP and TD. Our approach achieved a Dice of 0.92 and IoU of 0.89, with only 5.82$\%$ mislabeled vertices, indicating a $\sim$15$\%$ improvement (Fig ~\ref{fig2}A). To further assess our clustering strategy, we independently validated the RNMF-KMeans approach using features from the TReND framework. The results also demonstrate a $\sim$15$\%$ accuracy gain over standalone NMF, kPCA and ICA (Fig ~\ref{fig2}B). This validation highlights the robustness of our framework in preserving structural patterns while mitigating noise and spurious signals, a key challenge in neonatal fMRI data.
\begin{figure}
\includegraphics[width=\textwidth]{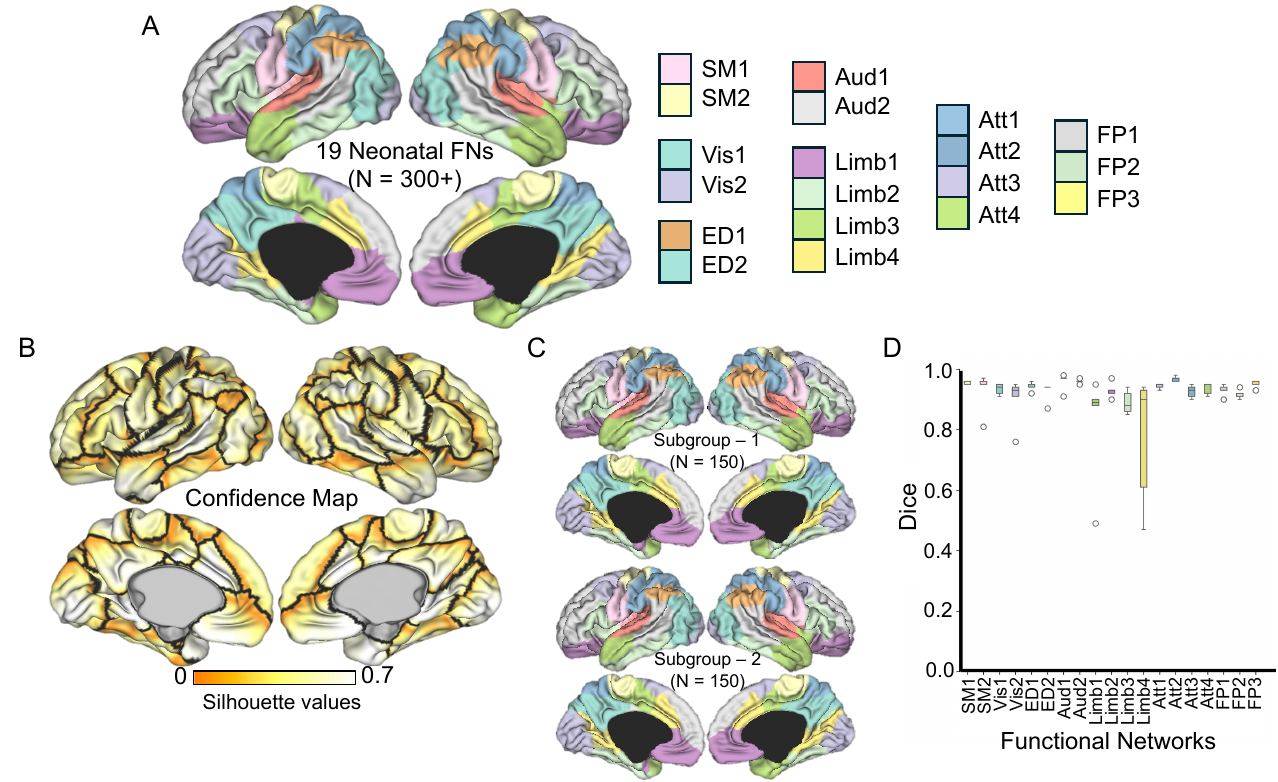}
\caption{Fine-grained parcellation of 19 functional networks in neonates derived from a 300-subject dHCP cohort. \textbf{A.} Neonatal cortical parcellation of 19 FNs. \textbf{B.} Confidence map representing the reliability of the 19 neonatal FNs parcellation. \textbf{C.} Discovery and replication of the 19 neonatal FNs cortical parcellation. \textbf{D.} Dice score evaluation of FN-wise via bootstrap analysis, assessing the consistency of the parcellation.} \label{fig4}
\end{figure}
\subsection{Coarse parcellation of 7 and 19 functional networks in neonates}
Building on the validation with the simulated dataset, we applied the TReND framework to the dHCP cohort. Our cluster stability analyses identify 7 and 19 FNs as optimal starting points for delineation of neonatal cortex (Fig \ref{fig3}A). The 7 FNs atlas provide a coarse parcellation of developing networks, derived from 300 term neonates in the dHCP cohort (Fig \ref{fig3}B). Confidence maps reveal lower confidence at boundary regions, suggesting further subdivision (Fig \ref{fig3}C). Bootstrap analysis with 100 combinations confirms high reproducibility, with the 7 FNs atlas achieving an average Dice of $\sim$0.92 (Fig \ref{fig3}D). Network-wise variability remains low (Standard deviation (SD) $\sim$0.05 Dice), except for the attention network, which exhibits slightly higher variation (Fig \ref{fig3}E). We identify emerging default network in neonates, where the absence of frontal and temporal components reflects the incomplete maturation of higher-order FNs compared to adults (Fig \ref{fig3}F) \cite{ref24}. In contrast, primary networks, such as somatomotor and visual FNs, exhibit great consistency with the adult atlas (Fig \ref{fig3}F). This finding highlights why adult parcellation is not transferable to neonates, as several higher-order FNs are yet to fully develop.

Similarly, the 19 FNs atlas provide a finer-grained parcellation (Fig \ref{fig4}A), further fractionating low-confidence boundary regions from the 7 FNs, reflecting the hierarchical organization of neonatal FC (Fig \ref{fig4}B). This atlas also achieves an average Dice of $\sim$0.90, maintaining strong reproducibility despite increased granularity (Fig \ref{fig4}C). Most networks exhibit an SD of $\sim$0.09 Dice, except for the limbic 4 network, which shows higher variability (Fig \ref{fig4}D).
\begin{figure} 
\includegraphics[width=\textwidth]{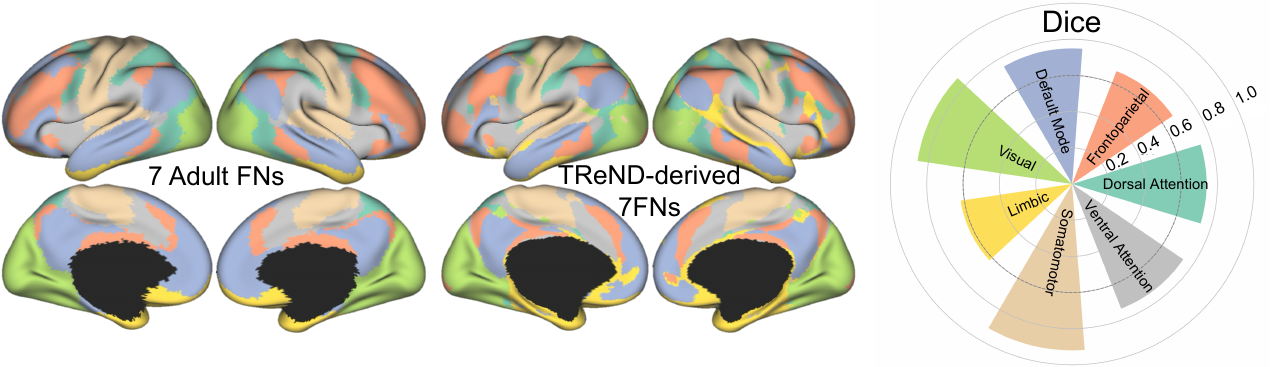} 
\caption{Validation of TReND's performance on the HCP-YA dataset. Consistency of the 7 FNs identified by TReND compared to the Yeo 7-network atlas evaluated with dice.} \label{fig5}
\end{figure}

\subsection{Cross-dataset performance of TReND using HCP-YA}
Although the TReND framework is developed for the neonatal cortex parcellation, we evaluated its applicability on the HCP-YA dataset using the Yeo 7-network atlas as ground truth \cite{ref3}. All networks, except the limbic network, achieved a Dice score of $\sim$0.7 or higher, with the somatomotor and visual networks reaching $\sim$0.9 (Fig \ref{fig5}). These variations may have arisen due to TReND's voxel-wise rather than ROI-wise approach and cohort differences. 
% Furthermore, its ability to identify neonatal FNs with a limited sample size of just 19 subjects highlights its practical applicability (Fig \ref{fig5}B).
\section{Conclusion}
Our study introduced TReND, a novel transformer-autoencoder framework with RNMF for neonatal FN parcellation. Given that adult FNs are not directly transferable to neonates due to their incomplete maturation, our approach provides a dedicated solution for neonatal brain functional organization. The resulting 7- and 19-network parcellations aim to establish standardized neonatal functional atlases, paving the way for future research in early brain development.

\begin{comment}  %% removed for anonymized MICCAI 2025 submission.
    
    % The following acknowledgement and disclaimer sections should be removed for the double-blind review process.  
    % If and when your paper is accepted, reinsert the acknowledgement and the disclaimer clause in your final camera-ready version.

\begin{credits}
\subsubsection{\ackname} A bold run-in heading in small font size at the end of the paper is
used for general acknowledgments, for example: This study was funded
by X (grant number Y).

\subsubsection{\discintname}
It is now necessary to declare any competing interests or to specifically
state that the authors have no competing interests. Please place the
statement with a bold run-in heading in small font size beneath the
(optional) acknowledgments\footnote{If EquinOCS, our proceedings submission
system, is used, then the disclaimer can be provided directly in the system.},
for example: The authors have no competing interests to declare that are
relevant to the content of this article. Or: Author A has received research
grants from Company W. Author B has received a speaker honorarium from
Company X and owns stock in Company Y. Author C is a member of committee Z.
\end{credits}

\end{comment}

\end{document}